**RESEARCH ARTICLE**

# Optimization of pulsed saturation transfer MR fingerprinting (ST MRF) acquisition using the Cramér-Rao bound and sequential quadratic programming

Nikita Vladimirov[1]  | Moritz Zaiss[2,3]  | Or Perlman[1,4]

[1]Department of Biomedical Engineering, Tel Aviv University, Tel Aviv, Israel

[2]Institute of Neuroradiology, University Hospital Erlangen, Friedrich-Alexander-Universität Erlangen-Nürnberg (FAU), Erlangen, Germany

[3]Department of Artificial Intelligence in Biomedical Engineering, Friedrich-Alexander-Universität Erlangen-Nürnberg (FAU), Erlangen, Germany

[4]Sagol School of Neuroscience, Tel Aviv University, Tel Aviv, Israel

**Correspondence**
Or Perlman
email: orperlman@tauex.tau.ac.il

**Funding Information**
This work was supported by the Ministry of Innovation, Science and Technology, Israel, and a grant from the Blavatnik Artificial Intelligence and Data Science Fund, Tel Aviv University Center for AI and Data Science (TAD). This project was funded by the European Union (ERC, Baby-Magnet, project no. 101115639).

**Purpose:** To develop a method for optimizing pulsed saturation transfer MR fingerprinting (ST MRF) acquisition.

**Methods:** The Cramér-Rao bound (CRB) for variance assessment was employed on Bloch-McConnell-based simulated signals, followed by a numerical sequential quadratic programming optimization and basin-hopping avoidance of local minima. Validation was performed using L-arginine phantoms and healthy human volunteers (n=4) at 3T while restricting the scan time to be less than 40 s.

**Results:** The proposed optimization approach resulted in a significantly improved agreement with reference gold standard values in vivo, compared to baseline non-optimized protocols (8% lower NRMSE, 7% higher SSIM, and 15% higher Pearson's r value, p<0.001).

**Conclusion:** The combination of the CRB with sequential quadratic programming and a rapid Bloch-McConnell simulator offers a means for optimizing and accelerating pulsed CEST and semisolid magnetization transfer (MT) MRF acquisition.

**KEYWORDS:**
CEST, MT, Optimization, MR Fingerprinting, Quantitative MRI

## 1 | INTRODUCTION

Chemical exchange saturation transfer (CEST) and semisolid magnetization transfer (MT) MRI leverage the saturation transfer (ST) mechanism to extract molecular information[1]. Over the years, these approaches have proven valuable for preclinical biological investigations and clinical human studies[2,3,4]. By informing on molecular events associated with altered metabolite composition, compound concentration, or pH, ST MRI provides a radiation-free alternative to positron emission tomography (PET) and single photon emission computed tomography (SPECT) while offering improved spatial resolution compared to MR spectroscopic imaging (MRSI)[5,6,7].

ST MRI operates by applying selective radio-frequency pulses to saturate exchangeable protons on mobile proteins, peptides, macromolecules, or metabolites, which then transfer this saturation to bulk water protons. The semisolid MT contrast is a known biomarker for myelin integrity in multiple sclerosis[8]. In addition, amide proton CEST has been employed to detect pH changes during stroke[9,10] and to differentiate tumor progression from radiation necrosis in glioma patients[11,12], while glutamate CEST has been used to characterize neurodegenerative, psychiatric, and oncological disease[13,14,15]. Many other ST-based applications have been reported and are the subject of research. These include glycogen imaging[16], protein aggregation detection[17], reporter gene and liposome imaging[18,19,20], glucose uptake analysis[21,22], and cardiac metabolism characterization[23].

While ST-*weighted* imaging has demonstrated a marked potential, especially for brain tumors[24], the technique still faces several challenges that must be overcome before the full extent of this contrast mechanism can be exploited. A key limitation is the semi-quantitative nature of CEST and semisolid MT imaging; the native ST signal not only contains the product of contributions from the proton volume fraction ($f_s$) and exchange rate ($k_{ex}$) of multiple compounds and metabolites, but is also affected by water $T_1$ and the particular parameters of the pulse sequence used[25].

Quantitative imaging is a desirable outcome when developing MR methods, because it enables reproducibility and cross-study comparisons while facilitating a physically meaningful interpretation of image data[26,27,28]. ST MRI is no exception, and accurate quantification of proton exchange parameters has been the goal of various previous research efforts[29,30,31]. Several foundational methods, such as quantifying exchange rates using the saturation time/saturation power (QUEST/QUESP)[2,29,32], Omega plots[33,34], and Bloch-McConnell (BM) fitting[2,35], rely on analytical models derived from the BM equations. These methods typically assume steady-state saturation and complete relaxation, leading to long scan times.

Magnetic resonance fingerprinting (MRF) is a different quantification paradigm that utilizes non-steady state, rapidly acquired data[36]. In MRF, a simulated dictionary of synthetic signal trajectories is generated and compared (e.g., using dot-product) to the experimentally acquired data. The best-matching dictionary entry is then used to determine the most suitable tissue parameter set for each voxel. Several years after its introduction for water relaxometry, MRF was adapted for

ST MRI[37,38,39,40,41,42,43]. While early ST MRF reports employed a pseudo-random acquisition protocol[37], later studies revealed that the encoding capability is heavily dependent on the pulse sequence parameters used[43,44]. As a result, intensive optimization is required in order to employ ST MRF in new applications or to shorten the scan time. Unfortunately, the extremely large size of the multi-proton-pool parameter space associated with ST MRI, makes an exhaustive search for the optimal pulse sequence parameters impractical[43]. Several deep-learning-based optimization strategies have recently been developed in order to address this challenge[45,46]. However, these approaches rely on the analytical solution of the Bloch-McConnell equations, which is not available for multi-pool *pulsed* ST acquisition, as commonly applied in clinical scanners. This lack prevents the use of ST MRF in a variety of practical human imaging applications.

The Cramer-Rao Bound (CRB), a statistical theoretic bound for parameter estimation variance[47], has been considered previously for the optimization of quantitative water $T_1$ and $T_2$ acquisition protocols[48]. Preliminary phantom studies have also indicated that it has the potential to reflect the discrimination ability of a given CEST MRF protocol[49].

Here, we describe a unified method for the automatic optimization of pulsed ST MRF acquisition protocols. Our protocol combines the CRB with a gradient-based iterative nonlinear programming algorithm[50] and a rapid numerical BM simulator[51]. While previous implementations of the CRB in the context of $T_1/T_2$ quantification used a state-space model to derive an analytical expression for the CRB[48], we now propose a numerical approach to address the problem posed by the lack of pulsed ST analytical solutions. The method was validated using phantoms and human volunteers in a clinical 3T scanner.

## 2 | METHODS

### 2.1 | ST MRF acquisition optimization pipeline

The acquisition protocol optimization pipeline comprises three main iterative building blocks: (1) Generating a dictionary of synthetic signals for the imaging scenario of interest using a rapid BM simulator. (2) Calculating the CRB, which represents the discrimination ability and expected variance across the entire dictionary, in the context of the proton exchange parameters to be quantified.

(3) Minimizing the CRB using a nonlinear programming algorithm that performs gradient-based steps in the acquisition parameter space.

**Generation of an ST MRF dictionary**

The suggested optimization process mandates the de-novo generation of a synthetic signal dictionary for any change in the acquisition parameter set. To overcome the traditionally long numerical simulation time associated with solving the Bloch–McConnell equations for a saturation pulse train, we used a Pulseq-CEST-based[52] C++ implemented simulator. Further acceleration and improved dictionary generation capabilities were achieved by modifying and expanding the source code using a Python interface for parallel execution[51].

The simulator implemented the numerical solution of the following equation:

$$\mathbf{M}(t + \Delta t) = (\mathbf{M}(t) + \mathbf{A}^{-1}\mathbf{C}) \cdot e^{\mathbf{A}\Delta t} - \mathbf{A}^{-1}\mathbf{C} \quad (1)$$

with $T_2^*$ relaxation simulated as described in [52,53]. Briefly, multiple signal trajectories of subvoxel isochromats were simulated with Cauchy-Lorentz distributed $\Delta B_0$ inhomogeneities and summed to obtain the final magnetization signal. Additional dictionary details can be found in Supporting Information Table S1.

**Proton exchange parameter quantification**

The commonly used dot-product metric was used to reconstruct the quantitative parameter maps[37,44]:

$$(\hat{f_s}, \hat{k_{sw}})_{i,j} = \underset{f_s, k_{sw}}{\arg\max} \frac{< \mathbf{e}_{i,j}^T, \mathbf{d}(f_s, k_{sw}) >}{||\mathbf{e}_{i,j}||_2 \cdot ||\mathbf{d}(f_s, k_{sw})||_2} \quad (2)$$

where $\mathbf{e}_{i,j}$ is the experimental signal trajectory at pixel $(i, j)$ and $\mathbf{d}(f_s, k_{sw})$ is the dictionary entry simulated signal that corresponds to a certain proton volume fraction $f_s$ and exchange rate $k_{sw}$ parameter pair.

**CRB-guided optimization**

The main goal of the optimization process is to identify a set of acquisition parameters that enable accurate quantification of the proton exchange parameters for a given (or minimal) scan time. In the context of the CRB, this requirement translates into efficient discrimination between different signal trajectories. The optimization was performed using the sequential quadratic programming (SQP) algorithm, an iterative approach that seeks the optimum of a constrained nonlinear problem[50]. To avoid local minima (a previously observed challenge for Cramér-Rao based optimization[48]), the SQP was further combined with a basin-hopping optimization strategy, which introduces random perturbations and "jumps" throughout the solution (acquisition parameter) space[54].

The estimation of the normalized Cramer Rao Bound[47,49] for each synthetic signal dictionary was

formulated as:

$$nCRB(\boldsymbol{\theta}) = \sqrt{I(\boldsymbol{\theta})^{-1}}/\boldsymbol{\theta} \quad (3)$$

$$I(\boldsymbol{\theta}) = -E\left[\frac{\partial^2 \ln(p(\boldsymbol{x}; \boldsymbol{\theta}))}{\partial \boldsymbol{\theta}^2}\right] \overset{\text{p} \simeq \mathbb{N}(\boldsymbol{s}; \sigma)}{=}$$
$$\frac{1}{\sigma^2} \sum_{n=0}^{N-1} \frac{\partial s[n; \boldsymbol{\theta}]}{\partial \boldsymbol{\theta}}^T \cdot \frac{\partial s[n; \boldsymbol{\theta}]}{\partial \boldsymbol{\theta}} \quad (4)$$

where $s[n]$ is an MRF signal trajectory simulated as the transverse part of the magnetization vector of water $s[n] = \sqrt{M_x^2 + M_y^2}$ at the end of the readout. The signal differential with respect to the quantification parameters $\frac{\partial s[n; \boldsymbol{\theta}]}{\partial \boldsymbol{\theta}}$ was calculated numerically as a two-point approximation on the MRF dictionary grid. The CRB was calculated with respect to $\boldsymbol{\theta} = (f_s, k_{sw})$ and constituted a 2×2 matrix. The optimization loss $\mathrm{L_{CRB}}$ was defined as the matrix trace:

$$\mathrm{L_{CRB}} = \mathrm{tr}(nCRB(\boldsymbol{\theta})) \quad (5)$$

$$\hat{\boldsymbol{\varphi}}_{acq} = \underset{\boldsymbol{\varphi}_{acq}}{\arg\min}(\mathrm{L_{CRB}}) \quad (6)$$

where $\hat{\boldsymbol{\varphi}}_{acq}$ is the acquisition parameter matrix, which can include the saturation pulse power vector $\boldsymbol{B_1}[\mathrm{n}]$, the frequency offset vector $\boldsymbol{\Delta\omega}[\mathrm{n}]$, and any other acquisition parameters.

The CRB-guided SQP pipeline was used to optimize two types of short ST MRF protocols: (i) L-arginine CEST phantom imaging with a fixed saturation pulse frequency offset (3 ppm) and varied saturation pulse powers (0-4 $\mu$T at 3T or 0-6 $\mu$T at 7T). (ii) Semisolid MT brain imaging at 3T, with varied saturation pulse powers (0-4 $\mu$T) and frequency offsets (10 - 75 ppm). The data acquisition time for both protocols was restricted to less than 40 s by setting the number of raw MRF images to four or eight, while using relatively short saturation and recovery times (see section 2.4). An initial saturation pulse pattern was randomly generated and fed into the optimization protocol, which yielded an optimized series of saturation pulse powers and frequency offsets. To assess the reproducibility and consistency of the optimization pipeline, the process was repeated at least four times for each imaging scenario and schedule length (Supporting Information Figures S1-S3) and compared across different subjects.

## 2.2 | Phantom preparation

CEST phantoms were prepared as previously described[37,44,55]. Briefly, L-arginine was suspended in PBS at 25, 50, 75, 100, and 200 mM and titrated with NaOH to various pH levels between 4 and 5.5. The different solutions were placed in 2-mL glass vials with sets of 3 vials placed into 50-mL Falcon tubes (suitable for a preclinical scanner) or six 10-mL vials placed inside a commercially available phantom holder (Gold Standard Phantoms, UK, model MultiSample-120E, suitable for a clinical scanner).

## 2.3 | Human subjects

The research protocol was approved by the Tel Aviv University Institutional Ethics Board (study no. 0007572-2) and the Chaim Sheba Medical Center Ethics Committee (0621-23-SMC). Four healthy male volunteers (age $24.75 \pm 2.3$ years) were recruited and signed an informed consent form.

## 2.4 | Data acquisition

**Preclinical continuous wave imaging at 7T**
The L-arginine phantoms were imaged using a preclinical 7T MRI scanner (Bruker, Germany) as a first "sanity check", designed to ensure that the proposed optimization approach could automatically improve the parameter discrimination ability in a controlled environment, and for later comparison. A versatile 4-, 8-, or 30-raw-images-long MRF protocol was realized using the open-source code described previously[51]. This comprises a continuous wave (CW) rectangular saturation pulse with saturation time ($T_{sat}$) = 3 s, frequency offset $\Delta\omega$ = 3 ppm, and recovery time ($T_{rec}$) = 1 s, followed by a single-slice SE-EPI readout with a flip angle (FA) = $60°$ and echo time (TE) = 20 ms[37]. The matrix size was $64 \times 64$, and the field of view (FOV) was $32 \times 32$ mm$^2$.

**Clinical pulsed wave imaging at 3T**
The clinical scanner experiments were conducted using a 3T MRI equipped with a 64-channel head coil (Prisma, Siemens Healthineers). All acquisition schedules were implemented using the Pulseq prototyping framework[56] and the open-source Pulseq-CEST sequence standard[52].

Since CW saturation is not feasible for most clinical scanners because of specific absorption rate (SAR) and physical instrumentation constraints[57], pulsed wave (PW) irradiation must be used instead. In this case, a spin-lock saturation pulse train consisting of $13 \times 100$ ms, 50% duty cycle, with a $T_{rec}$ = 1 s was used[55]. The saturation block was fused with the 3D centric reordered snapshot EPI readout module described by Mueller et al.[58], providing a whole brain 1.81 mm isotropic resolution, with a FOV of $210 \times 210 \times 160$ mm$^3$, a flip angle of $12°$, and TE = 7.8 ms.

Previously established ST MRF protocols, were acquired as a reference gold standard[40,55] (Supporting Information Fig. S4). These protocols used the same acquisition parameters described above, but employed a different set of saturation pulse powers and frequency offsets, and relied on a larger number of raw images (30 instead of 4 or 8).

## 2.5 | Performance evaluation and statistical analysis

CRB-optimized L-arginine concentration maps were compared to known concentrations using the mean absolute percentage error (MAPE) metric. Optimized in vitro proton exchange rate maps were compared to steady-state quantification of the exchange rate QUESP-derived values[32], as described previously[37].

In the absence of absolute ground truth in vivo, the CRB-optimized proton volume fraction and exchange rate maps were compared to maps obtained using a reference gold standard MRF protocol[40]. Quantification performance in vivo was estimated using the normalized root mean squared error (NRMSE) metric:

$$\text{NRMSE}(I_{GS}, I_S) = \frac{\text{RMSE}(I_{GS}, I_S)}{\max(I_{GS}) - \min(I_{GS})}, \quad (7)$$

Where $I_{GS}$ is the gold standard reference map and $I_S$ is a map quantified using a proposed (CRB-optimized) sequence. We also calculated the structural similarity index metric (SSIM) and the Pearson correlation coefficient.

Statistical analysis used a two-tailed paired t-test, implemented with the open-source SciPy scientific computing library for Python[59] and presented as box plots. Statistics in the text are presented as mean $\pm$ SD. Differences were considered significant at p<0.05. Asterisk notations were designated as *p<0.05, **p<0.01, and ***p<0.001.

# 3 | RESULTS

## 3.1 | CW phantom imaging at 7T

Two representative comparisons of the L-arginine concentration and proton exchange rate maps obtained using the randomly generated and CRB-optimized acquisition protocols at 7T are shown in Figure 1. Additional preclinical parameter maps are available in Supplementary Information Figure S5. After optimization, MAPE dropped from from $85.3 \pm 71.2\%$ to $23.2 \pm 16.5\%$ for L-arginine concentration mapping (p<0.05) and from $30.4 \pm 9.3\%$ to $17.9 \pm 5.5\%$ for proton exchange rate mapping (p<0.001, Figure 2). The acquisition times for the optimized protocols were 16 s and 32 s for acquiring four or eight raw MRF images, respectively.

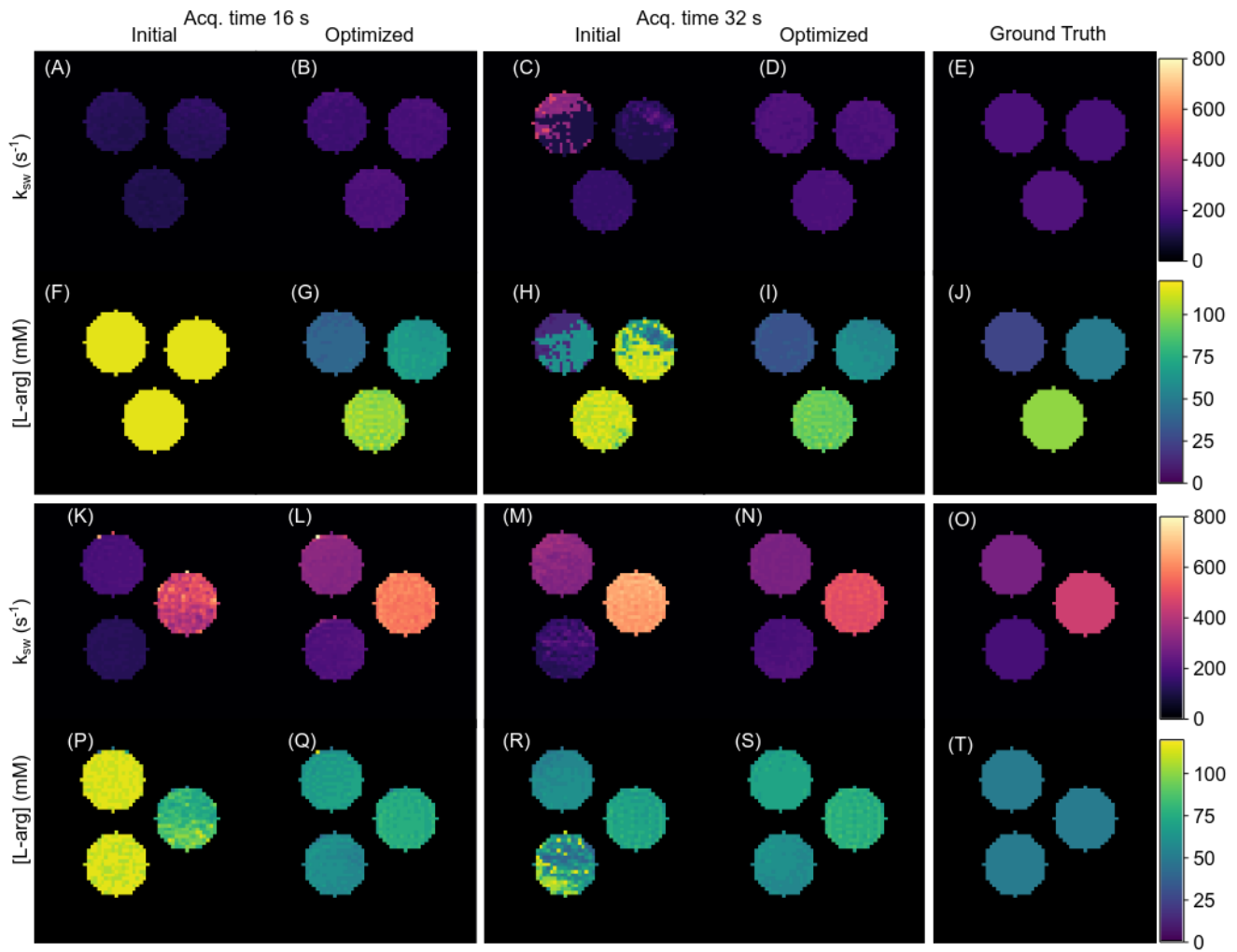

**FIGURE 1.** CW phantom imaging at 7T. (**Left**). Representative two pairs of pre (**A, F, K, P, C, H, M, R**) and post(**B, G, L, Q, D, I, N ,S**) CRB optimization L-arginine concentration and proton exchange maps, for an MRF schedule that acquires 4 or 8 raw images (in 16 s and 32 s, respectively). (**Right**). Ground truth concentrations (**J, T**) and QUESP-derived (**E, O**) gold standard exchange rates[37].

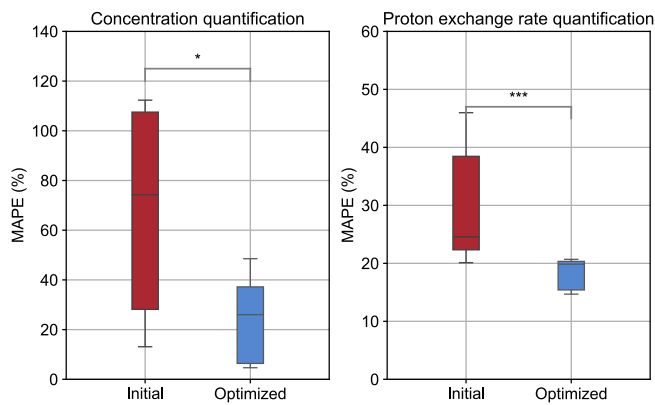

**FIGURE 2.** Statistical analysis of the mean absolute percent error (MAPE) across different L-arginine vials imaged at 7T. *p<0.05; ***p<0.001.

## 3.2 | PW phantom imaging using a clinical 3T scanner

A representative comparison of L-arginine concentration and proton exchange rate maps obtained using randomly generated and CRB-optimized PW acquisition protocols at 3T is shown in Figure 3. All other examples are available in Supporting Information Fig. S6. The MAPE for the proton exchange rate estimation decreased from 34.8±8.5% to 21.4±11.1% (p<0.05, Figure 4). Although the MAPE for estimating the L-arginine concentration decreased from 21.5±9.5% to 18.8±4.7%, the effect was not significant (p = 0.34). The acquisition time was 19.1 s and 38.2 s, for the MRF protocols that acquired four and eight raw images, respectively.

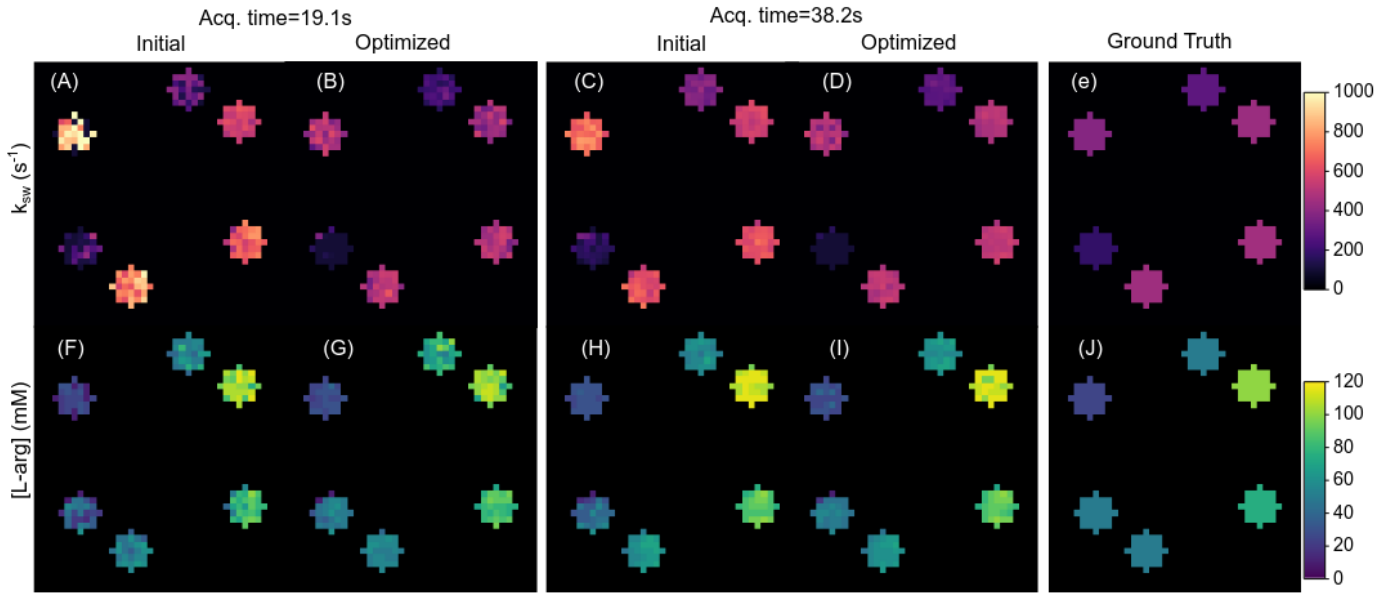

**FIGURE 3**. PW phantom imaging at 3T. (**Left**). A representative pair of pre (**A, F, C, H**) and post (**B, G, D, I**) CRB-optimization L-arginine proton exchange rate (**top**) and concentration (**bottom**) maps for an MRF schedule that acquires 4 or 8 raw images (in 19.1 s or 38.2 s, respectively). (**Right**). QUESP-derived gold standard exchange rates[37] (**E**) and ground truth concentrations maps (**J**).

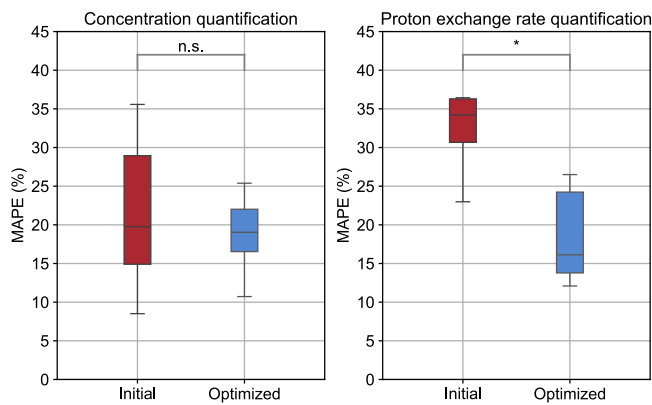

**FIGURE 4**. Statistical analysis of the mean absolute percent error (MAPE) across different L-arginine vials imaged using a PW MRF pulse sequence at a 3T clinical scanner. *p < 0.05.

## 3.3 | PW human brain imaging at 3T

Four pairs of pre- and post-CRB optimization parameter maps from a representative subject are shown in Figure 5. All the in vivo parameter maps are available in Supporting Information Figures S7-S10. In general, the optimization pipeline provided an improved signal-to-noise ratio (SNR) and produced images that were less noisy and more visually similar to the reference ground truth maps obtained using a previously established (and longer) acquisition protocol[40,55] (Figure 5C,F). A statistical analysis based on all optimization attempts, subjects, and slice images is provided in Figure 6. The CRB-guided semisolid MT volume fraction ($f_{ss}$) maps demonstrated better correlation with the ground truth reference than the maps obtained using randomly generated protocols (Pearson's r = 0.79±0.03 compared to 0.64±0.04), a higher SSIM (0.87±0.04 compared to 0.80±0.06), and a lower NRMSE (12±1% compared to 20±2%). Similarly, the proton exchange rate ($k_{ssw}$) maps obtained using the CRB-optimized protocols demonstrated higher SSIM (0.70±0.08 compared to 0.63±0.10), increased Pearson's correlation (0.41±0.09 compared to 0.26±0.09), and a lower NRMSE (20±5% compared to 28±8%). Importantly, all the improvements in metric performance described above (for both $f_{ss}$ and $k_{ssw}$) were statistically significant for all subjects (p<0.001, Figure 6). The acquisition time was identical to the PW phantom case, namely 19.1 s and 38.2 s for the MRF protocols that acquired four and eight raw images, respectively.

## 4 | DISCUSSION

ST MRF is an increasingly investigated method for the quantification of molecular processes in vivo. While the first studies randomly varied the saturation pulse parameters[37,39,38], subsequent investigations have demonstrated that the discrimination ability of the pulse sequence, and consequently, the quantification accuracy, are critically influenced by the specific acquisition parameters used[44,46,60,45]. Several gradient-descent-based approaches have been proposed for automatic optimization of the pulse sequence[43]. Such methods rely on a computational-graph-based formulation of the analytical BM solution, which exists for CW saturation and is well suited for preclinical animal studies[45]. However, the lack of accurate analytical solutions for saturation pulse trains hinders the optimization of ST MRF for human imaging (which typically utilizes PW acquisition to fulfill SAR requirements). The approach described here is designed to resolve this issue by using the CRB to guide a nonlinear iterative optimization that leverages a rapid numerical BM simulator.

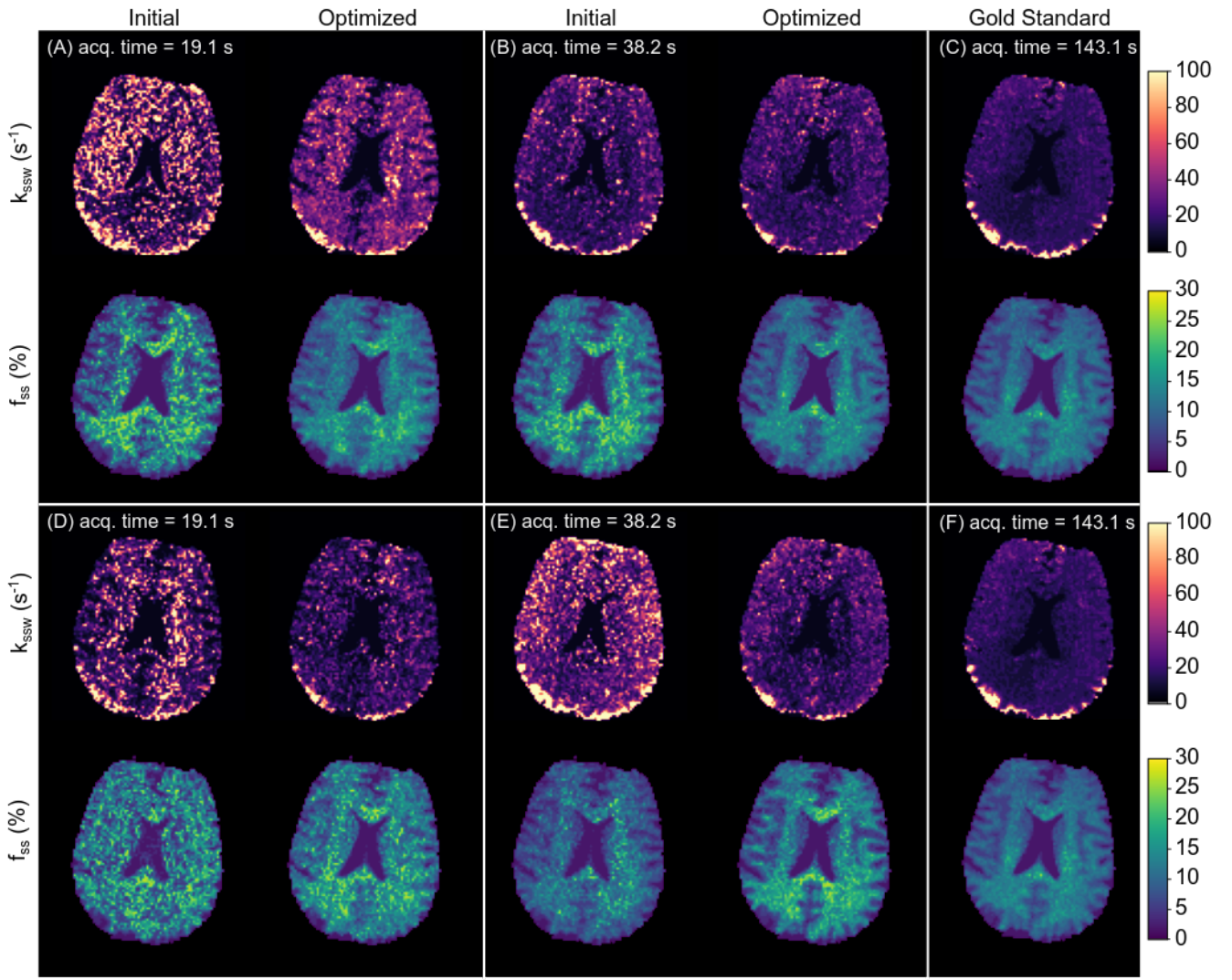

**FIGURE 5**. PW ST MRF imaging in a healthy human volunteer at 3T. **(A,B,D,E)** Pre- and post-CRB-optimization parameter maps from four independent procedures, each initiated using a randomly generated acquisition protocol (Supplementary Information Figure S3). **(C,F)** Gold standard reference maps obtained using a previously established (and longer) acquisition protocol[40,55].

An initial sanity check with CW-scanned phantoms revealed that applying the optimization pipeline consistently improved quantification accuracy (Figure 1, Figure 2, and Supporting Information Fig. S5). Similarly, a phantom experiment under PW conditions in a clinical 3T scanner provided better encoding capabilities following optimization (Figure 3). Notably, most of the improvement achieved was a result of more accurate quantification of the proton exchange rate (Figure 4), since the L-arginine concentrations maps for the randomly generated protocols were already in reasonable agreement with the ground truth (Figure 3, bottom panel).

In vivo, there was a marked improvement in visual similarity between the reconstructed quantitative parameter maps and the reference gold standard, following

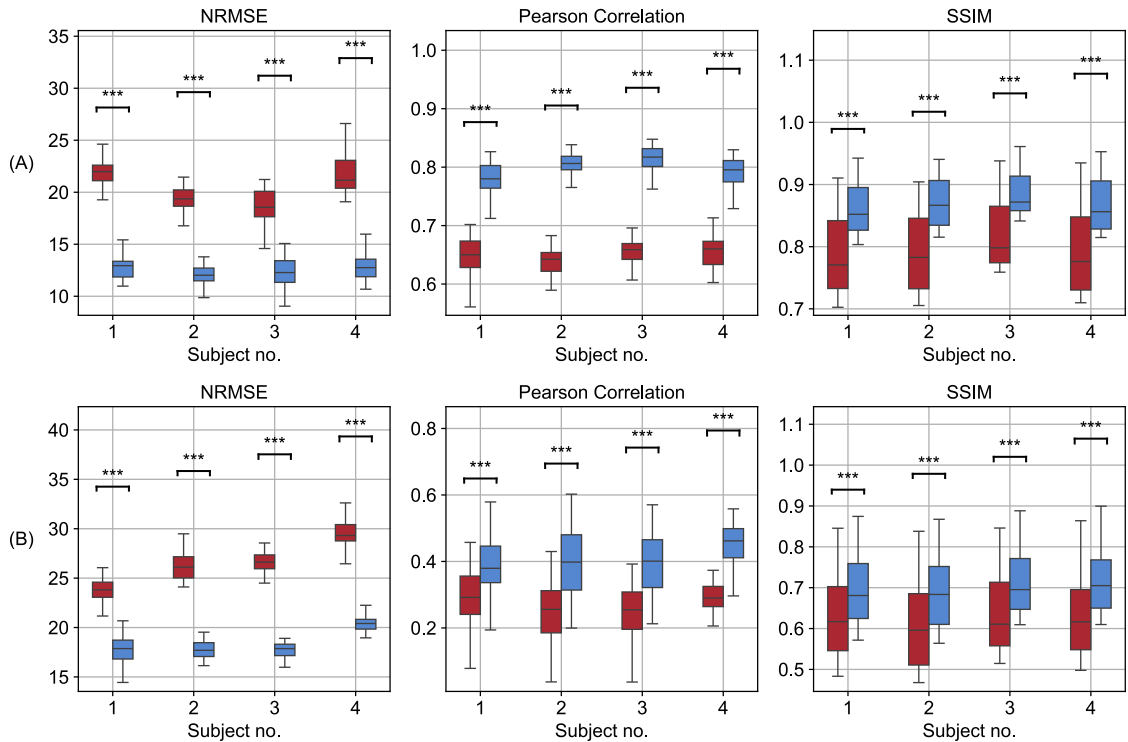

**FIGURE 6.** Statistical analysis of the in vivo semisolid MT proton volume fraction (**A**) and proton exchange rate (**B**) quantification performance. The NRMSE, SSIM, and Pearson's correlation values were calculated with respect to gold standard reference maps obtained using a previously established (and longer) acquisition protocol[40,55]. The red and blue box plots represent the initial and CRB-optimized acquisition protocols, respectively.

CRB-based optimization (Figure 5). A group analysis revealed significant improvements in the NRMSE, SSIM, and Pearson's r values across all subjects (p < 0.001, Figure 6). The improvement was most visually discernible for the very short acquisition protocols (scan time = 19.1 s, Figure 5A,D), where the parameter maps obtained for the randomly generated protocols were very noisy.

To gain a basic intuition of the decisions made by the optimizer, we performed a meta-analysis comparing the distribution of the saturation pulse parameters used by the randomly initialized and the CRB-optimized acquisition schedules (Supplementary Information Figure S11). The standard deviation of the optimized saturation pulse powers was significantly increased compared to the baseline in all cases (p<0.05). This can be reasoned by an improved encoding capability associated with using a wider range of powers, facilitating a more efficient saturation in various proton exchange rates. However, the variance of the saturation pulse frequency offset was not significantly higher following optimization, which can be explained by the broad spectral linewidth associated with the semisolid MT proton, which creates various "opportunities" for sufficient encoding, regardless of spanning a vast frequency offset range. The change in the mean saturation pulse parameters following optimization did not present a clear and consistent trend across all imaging cases. The changes in mean saturation parameter value were only significant for the CW saturation pulse power optimization (p<0.001).

The main challenges in interpreting PW MRF in clinical scanners compared to CW MRF in preclinical scanners are the reduced saturation efficiency (due to the lower duty cycle), the more difficult modeling of the saturation pulse train, and the larger field inhomogeneities. Although the field heterogeneity issue was mitigated here by simulating multiple subvoxel isochromats with Cauchy-Lorentz distributed $\Delta B_0$ inhomogeneities, future work could explicitly input experimentally measured field inhomogeneities in the per-voxel quantification[40].

This study deliberately used a very short (less than 40 s) pulse sequence, which generates only four or eight raw MRF images. The rationale was to explore the limits of ST MRF acceleration and potentially discover an optimized acquisition schedule for rapid and quantitative preliminary screening of ST effects using clinical scanners. While the post-optimization performance metrics were adequate for semisolid MT volume fraction mapping (Figure 6, top), the proton exchange rate quantification yielded a lower correlation with the reference maps obtained using a previously established and longer MRF protocol (30 raw images). This can be attributed to the well-known challenge of accurately quantifying the noisy semisolid MT proton exchange rate, where relatively subtle changes are spatially manifested across the brain[61,43].

The numerical nature of the proposed approach makes it suitable for a variety of pulse shapes, and any number of proton pools, with the primary cost being the optimization time. The CRB-SQP process took between 5 to 31 hours on a single desktop, depending on the imaging scenario (Supplementary Table S2). In this context, recent studies have shown that multi-pool CEST imaging necessitates the serial acquisition of $T_1$, $T_2$, semisolid MT and CEST encoded data and subsequent integration in a quantification scheme that combines the entire information[40,42]. Although we focused here on in-vivo semisolid MT quantification and a proof-of-concept in-vitro CEST imaging, future research will be needed to validate the use of the CRB-based optimization in more complex imaging scenarios (and pathologies).

An important factor in the success of the in-silico pre-scan offline optimization was the rapid BM simulator used, which not only incorporated C++ and Python backend parallelization for computational efficiency[51], but was also fully compatible with Pulseq-CEST standards. This enabled an accurate realization of the same PW saturation properties, as played out at the scanner[52].

This proof of concept study optimized two main acquisition parameters (Supporting Information Figures S1-S3): the saturation pulse power ($B_1$) and the frequency offset ($\Delta\omega$). However, the proposed pipeline can be readily extended to optimize additional parameters, such as the saturation pulse length, flip angle, and recovery time.

Another possibility for improvement is that here, we employed the classical dot product between the experimentally measured trajectories and the simulated dictionary entries as the quantification metric[36]. While the dot-product is often used and is simple to implement, it is possible that alternative NN-based quantification approaches[40,41,61,62] could provide improved accuracy. Interestingly, a recent study has shown that the CRB can be further utilized to improve such NN reconstruction by participating in the loss function normalization[63].

## 5 | CONCLUSION

The CRB-based optimization framework demonstrates the ability to accelerate 3D acquisitions of semisolid MT and CEST mapping by 3.75 to 7.5 fold, with a better performance than pseudo-randomly generated protocols. The unlocked optimization ability for pulsed saturation creates new opportunities to support future clinical ST research.

## ACKNOWLEDGMENTS

The authors thank Tony Stöcker and Rüdiger Stirnberg for their help with the 3D EPI readout. The study was supported in part by MOONSHOT-MED, a joint grant program from Tel Aviv University and Clalit Innovation, the innovation arm of Clalit Health Services, and the Edmond J. Safra Center for Bioinformatics at Tel Aviv University. This work was supported by the Ministry of Innovation, Science and Technology, Israel, and a grant from the Blavatnik Artificial Intelligence and Data Science Fund, Tel Aviv University Center for AI and Data Science (TAD). This project was funded by the European Union (ERC, BabyMagnet, project no. 101115639). Views and opinions expressed are, however, those of the authors only and do not necessarily reflect those of the European Union or the European Research Council. Neither the European Union nor the granting authority can be held responsible for them.

## Data and Code Availability Statement

The source code, phantom data, and 2D human sample data will become publicly available upon acceptance at https://github.com/momentum-laboratory/crb-optim. The reference gold standard L-arginine and semisolid MT MRF pulse sequences are available at the Pulseq online library[52,64]. The preclinical and all CRB-optimized pulse sequences can be reproduced using the open-source software described previously[51] and the parameters defined in Supplementary Information Figures S1-S3.

## Conflict of interest

The authors declare no potential conflict of interests.

## ORCID

*Nikita Vladimirov*  0000-0003-1943-9139
*Moritz Zaiss*  0000-0001-9780-3616
*Or Perlman*  0000-0002-3566-569X

# Supporting Information

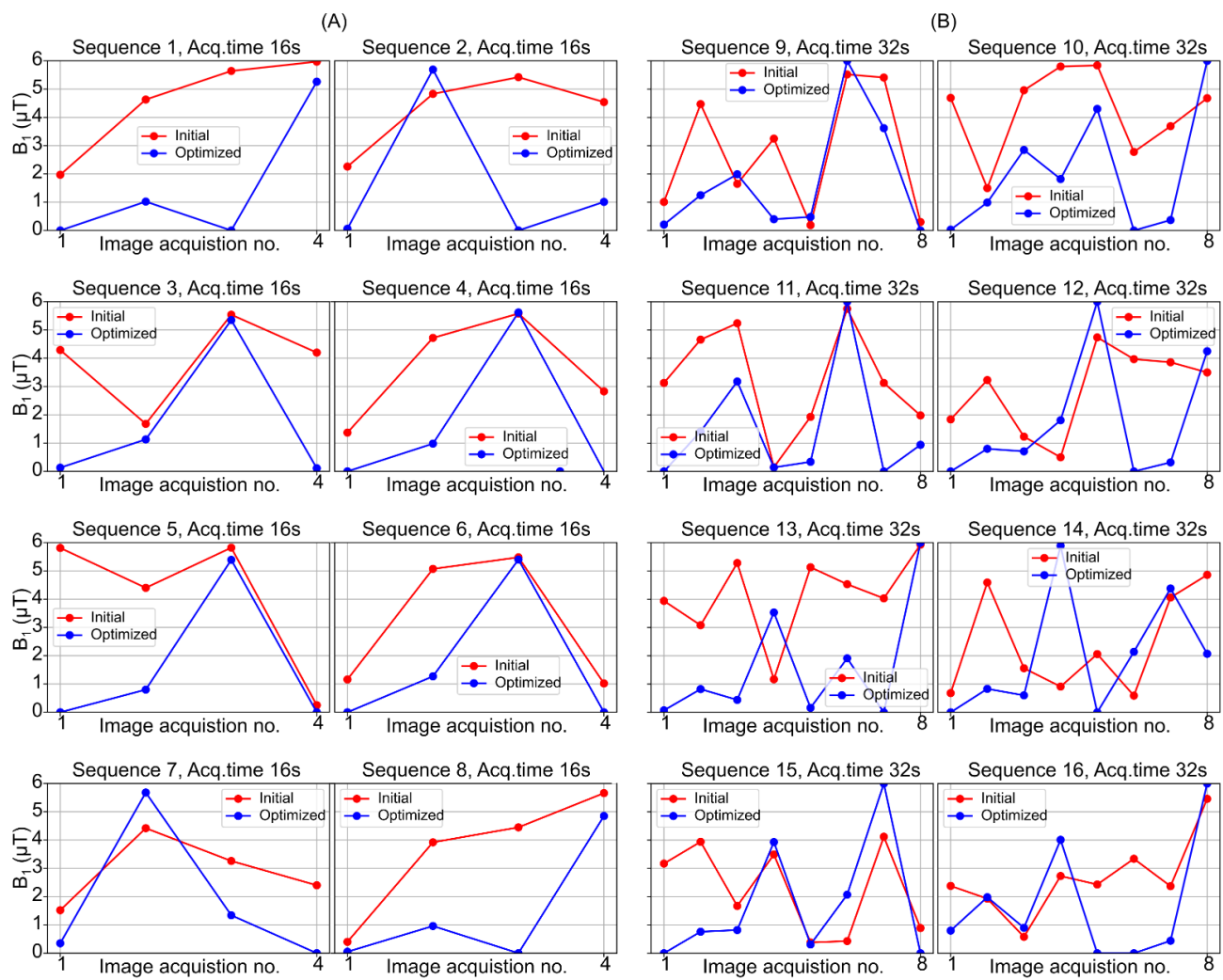

**Figure S1. Initial and optimized acquisition schedules for CW L-arginine phantom imaging at a preclinical 7T scanner**. (**A**) Schedules aimed to acquire four raw MRF images (acquisition time = 16 seconds). (**B**) Schedules aimed to acquire eight raw MRF images (acquisition time = 32 seconds).

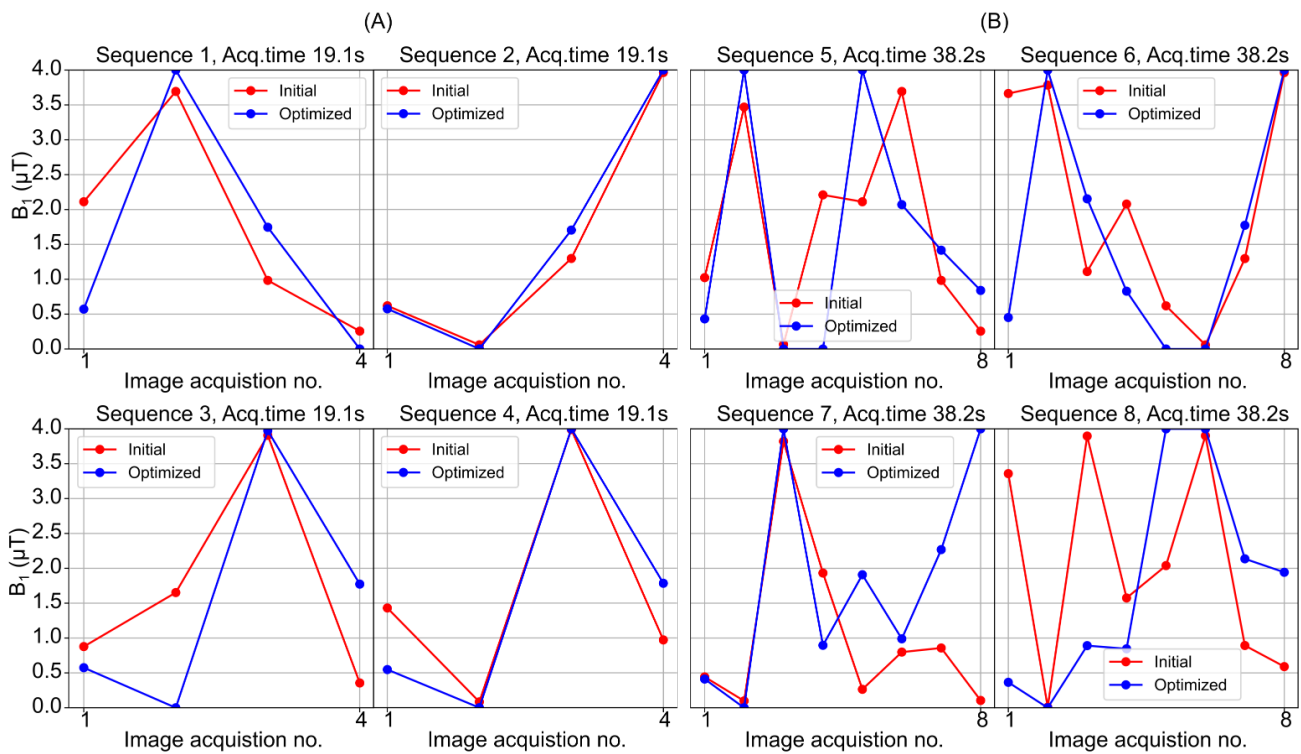

**Figure S2. Initial and optimized acquisition schedules for PW L-arginine phantom imaging at a clinical 3T scanner**. (**A**) Schedules aimed to acquire four raw MRF images (acquisition time = 19.1 seconds). (**B**) Schedules aimed to acquire eight raw MRF images (acquisition time = 38.2 seconds).

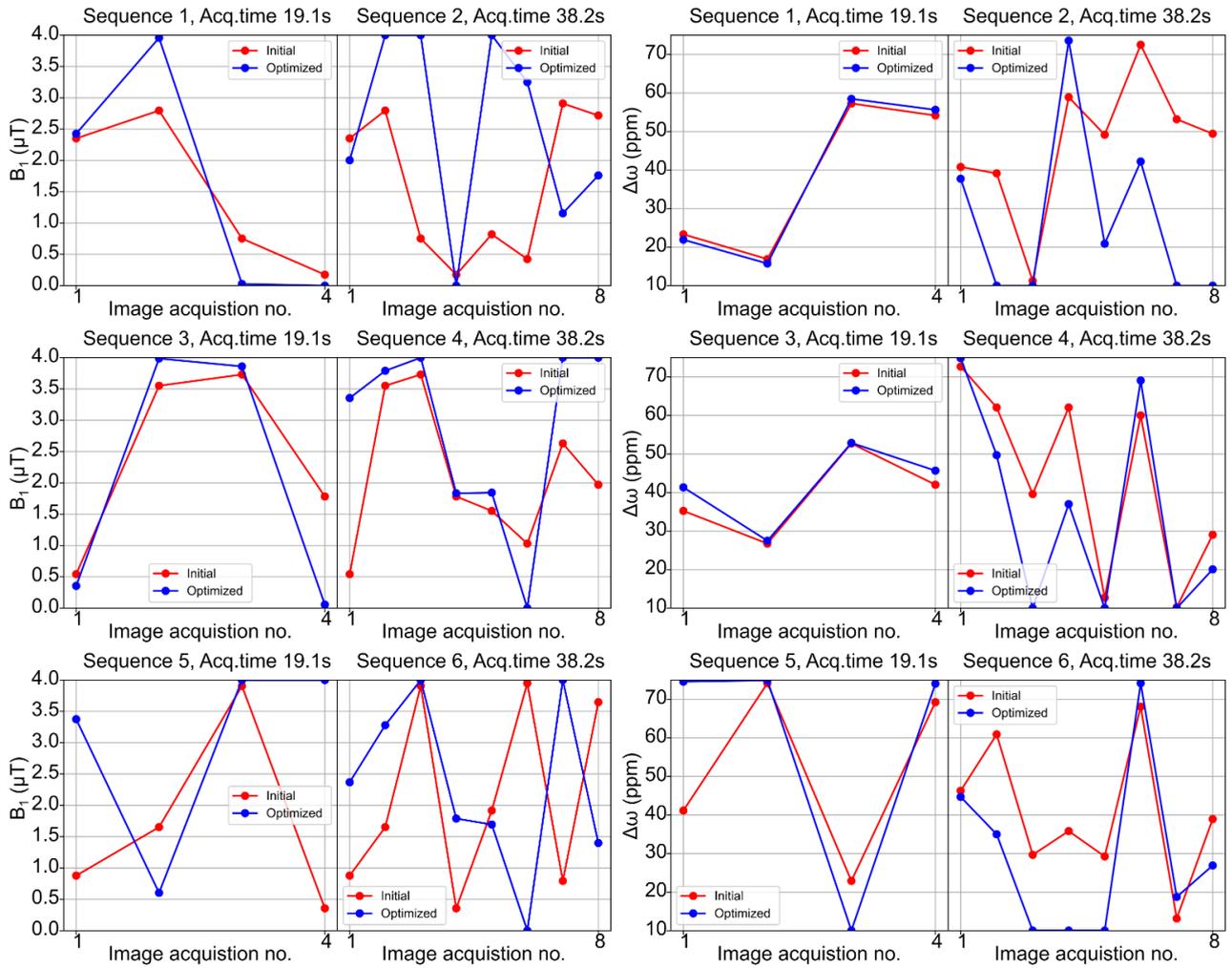

**Figure S3.** Initial and optimized acquisition schedules for PW in vivo human imaging at a 3T scanner, where both the saturation pulse power (B₁, left) and frequncy offset (Δw, right) were optimized.

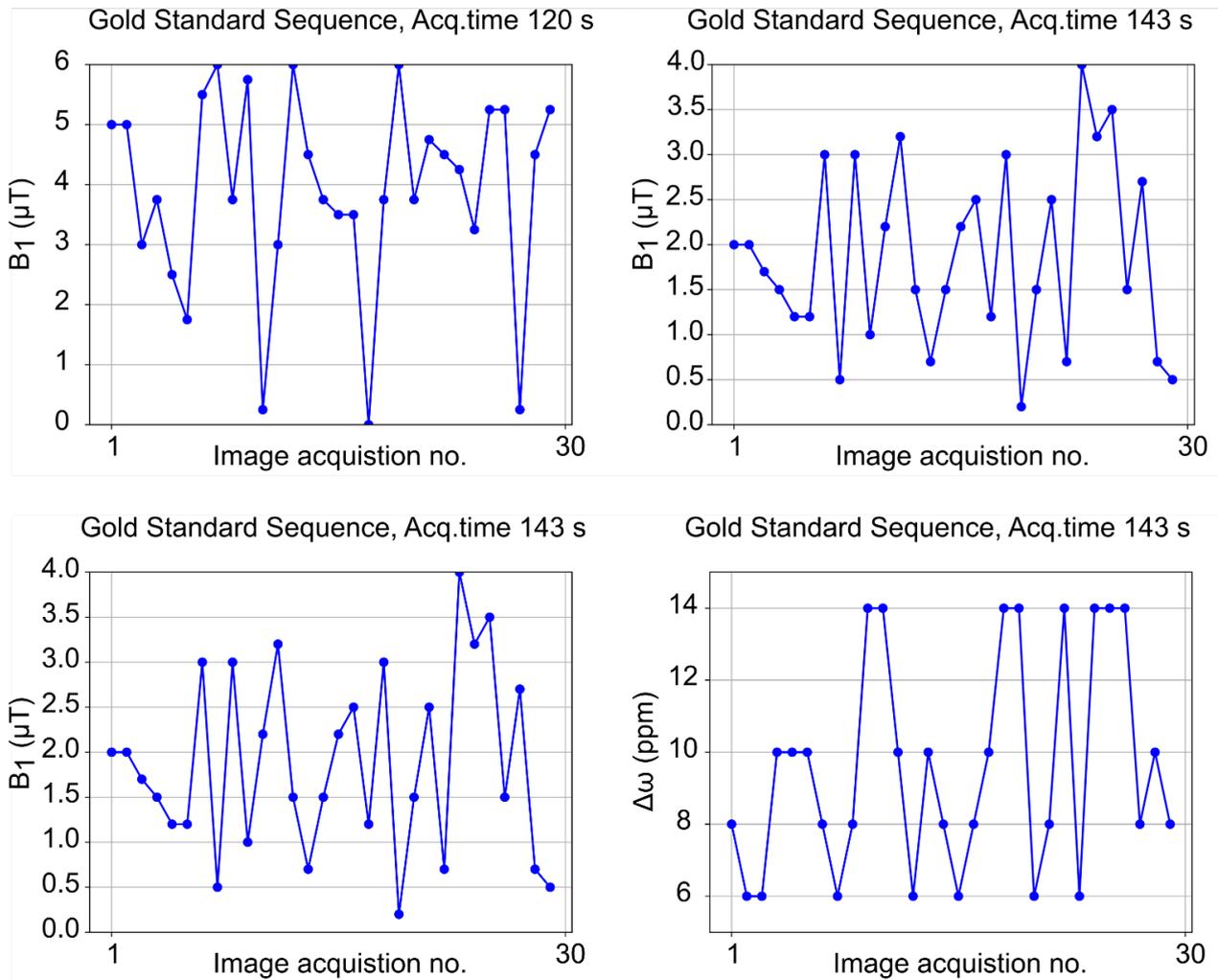

**Figure S4. Gold standard MRF acquisition protocols. (Top left)**. The saturation pulse powers used for CW phantom imaging in previously established reports[37]. A total of 30 images were acquired with a total acquisition time = 120 s. **(Top right)**. The saturation pulse powers used for PW phantom imaging[40]. A total of 30 images were acquired with a total acquisition time = 143 s). **(Bottom left)**. The saturation pulse powers and (**Bottom right**) frequency offsets used for PW human imaging[40]. A total of 30 images were acquired with a total acquisition time = 143 s. All other acquisition parameters are described in section 2.4.

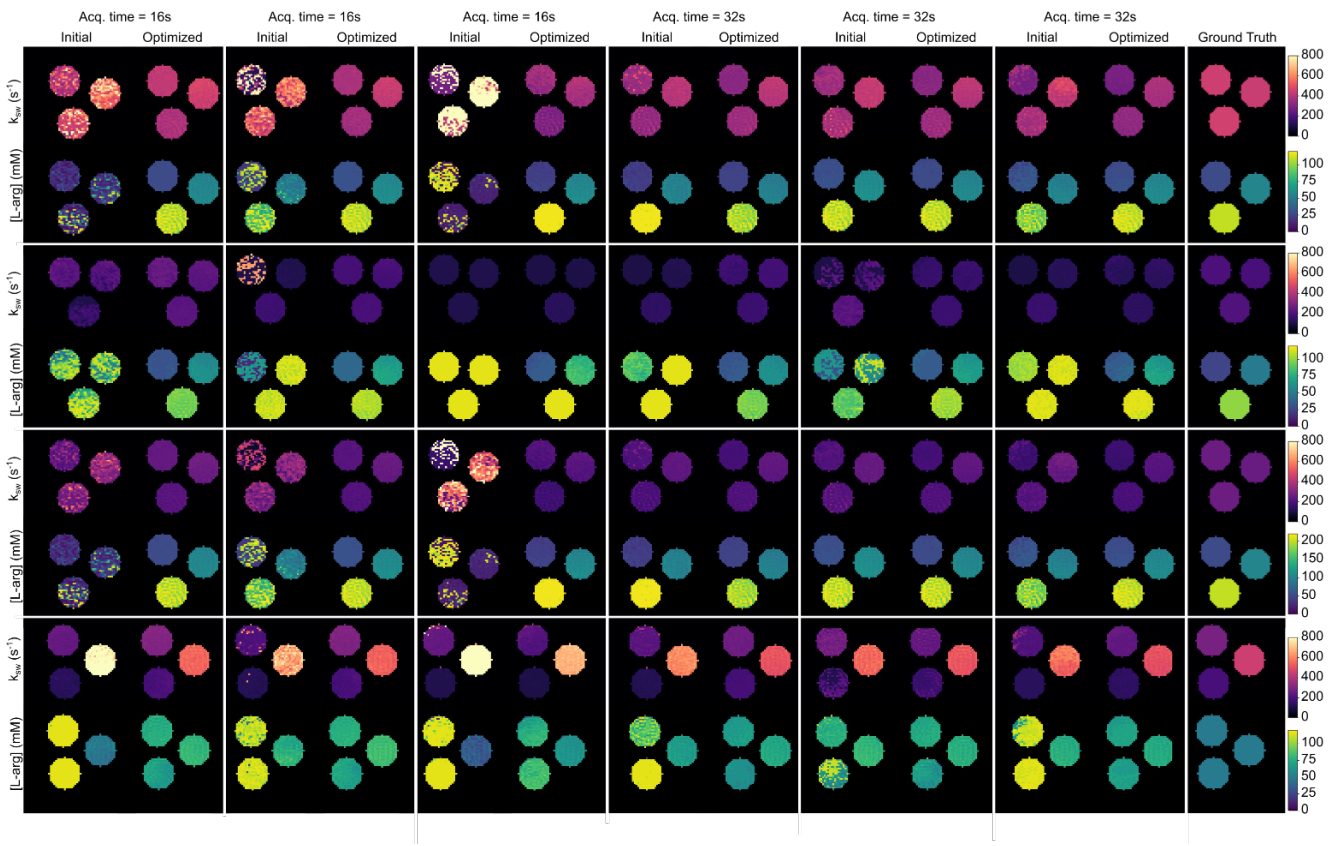

**Figure S5. Six additional examples for CW phantom imaging at 7T, before and following CRB-based optimization.** Each column contains the quantification maps obtained by applying the same initial and optimized pulse sequence pair across four different phantoms.

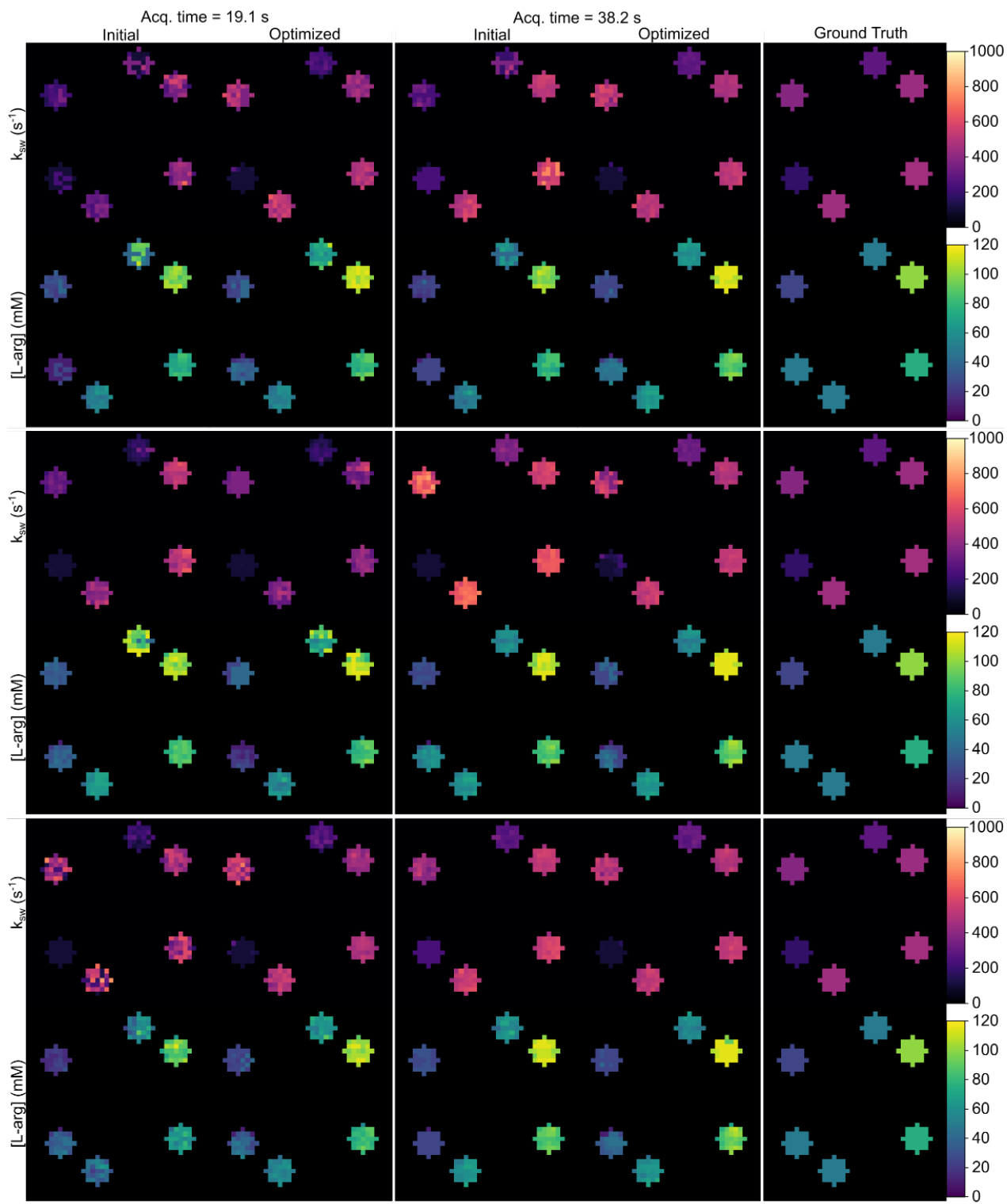

**Figure S6. The initial and optimized output parameter maps following all six optimization procedures performed for PW phantom imaging at 3T.**

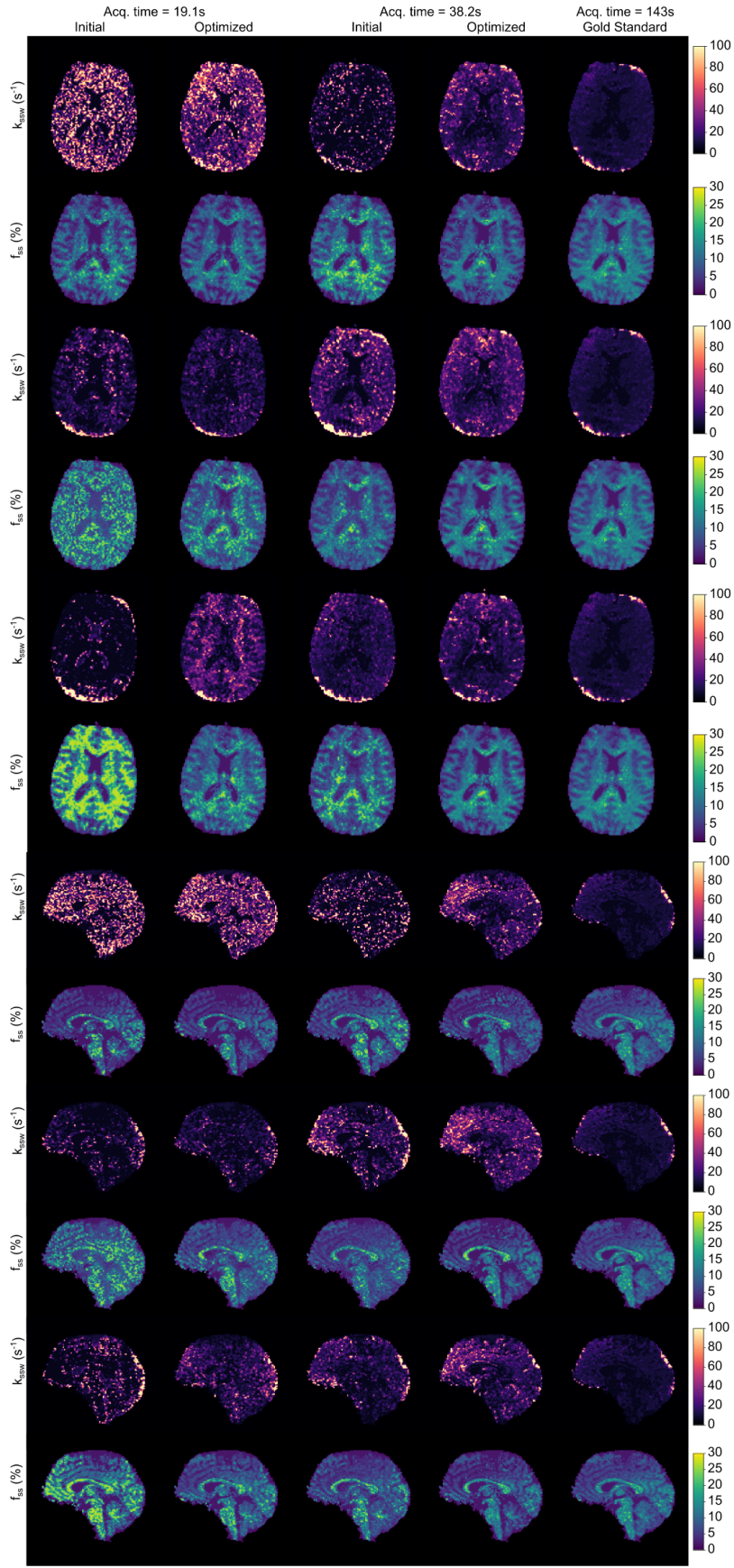

**Figure S7. PW ST MRF imaging in healthy human volunteer #1**. A representative slice is shown for each of the twelve optimization procedures performed (see Figure S3).

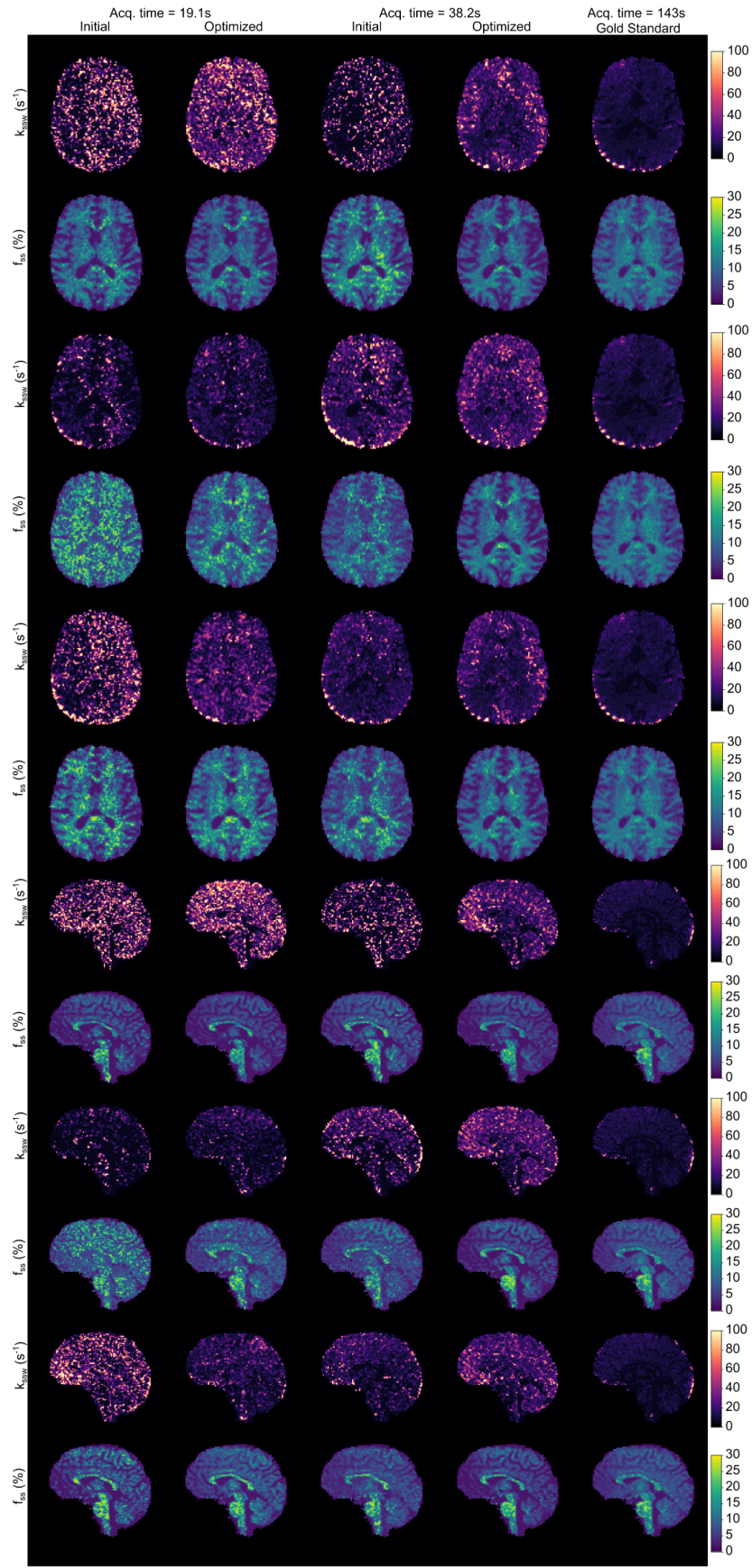

**Figure S8. PW ST MRF imaging in healthy human volunteer #2**. A representative slice is shown for each of the twelve optimization procedures performed (see Figure S3).

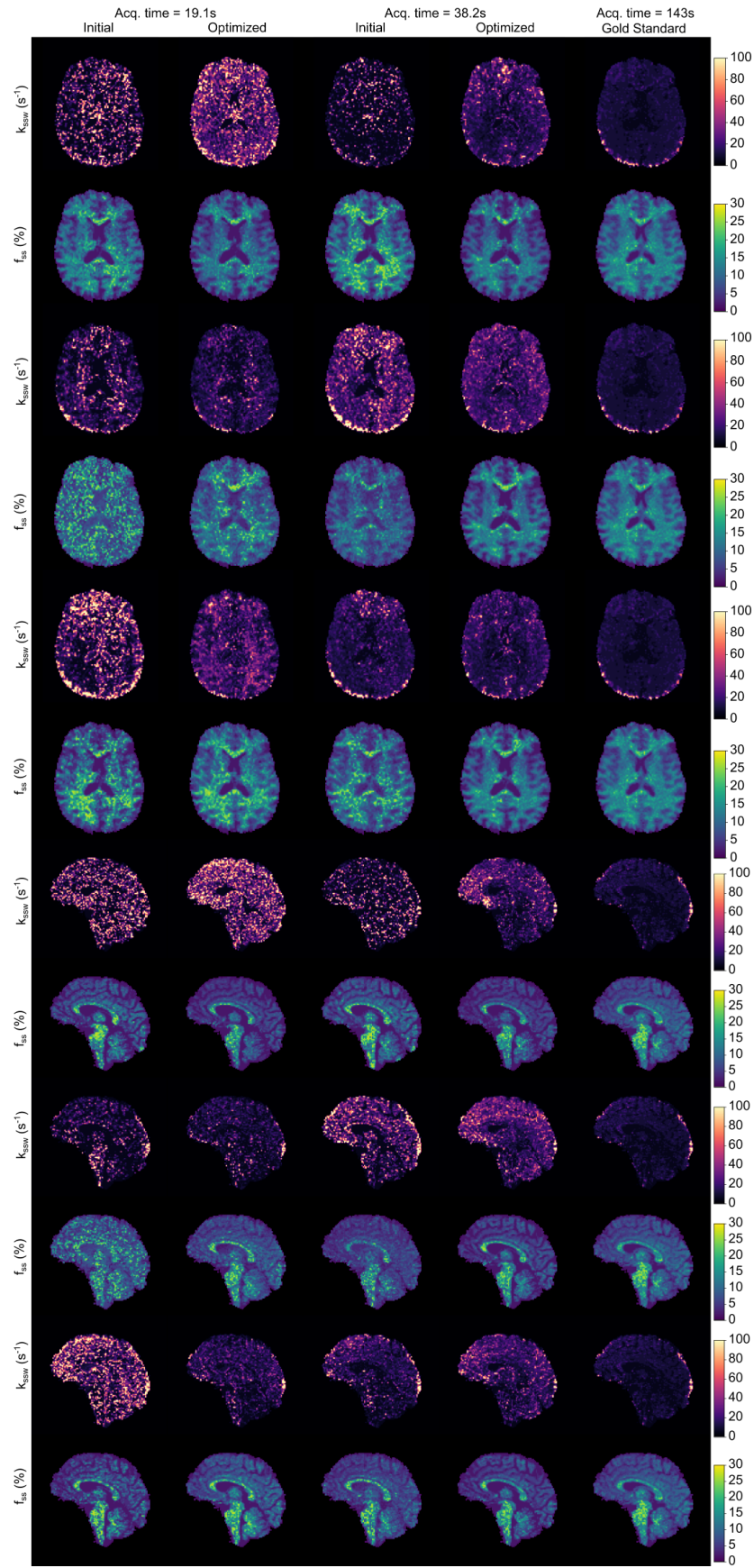

**Figure S9. PW ST MRF imaging in healthy human volunteer #3**. A representative slice is shown for each of the twelve optimization procedures performed (see Figure S3).

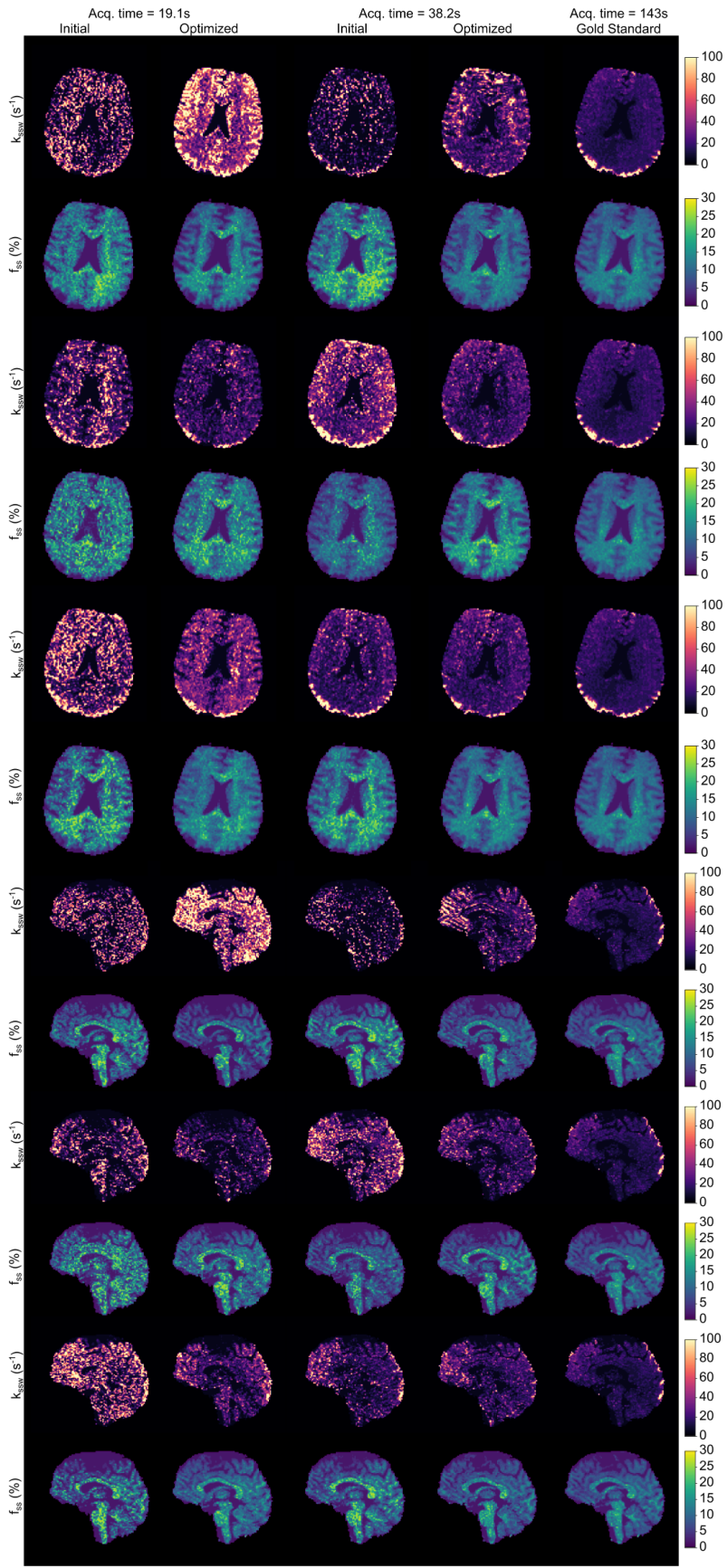

**Figure S10. PW ST MRF imaging in healthy human volunteer #4**. A representative slice is shown for each of the twelve optimization procedures performed (see Figure S3).

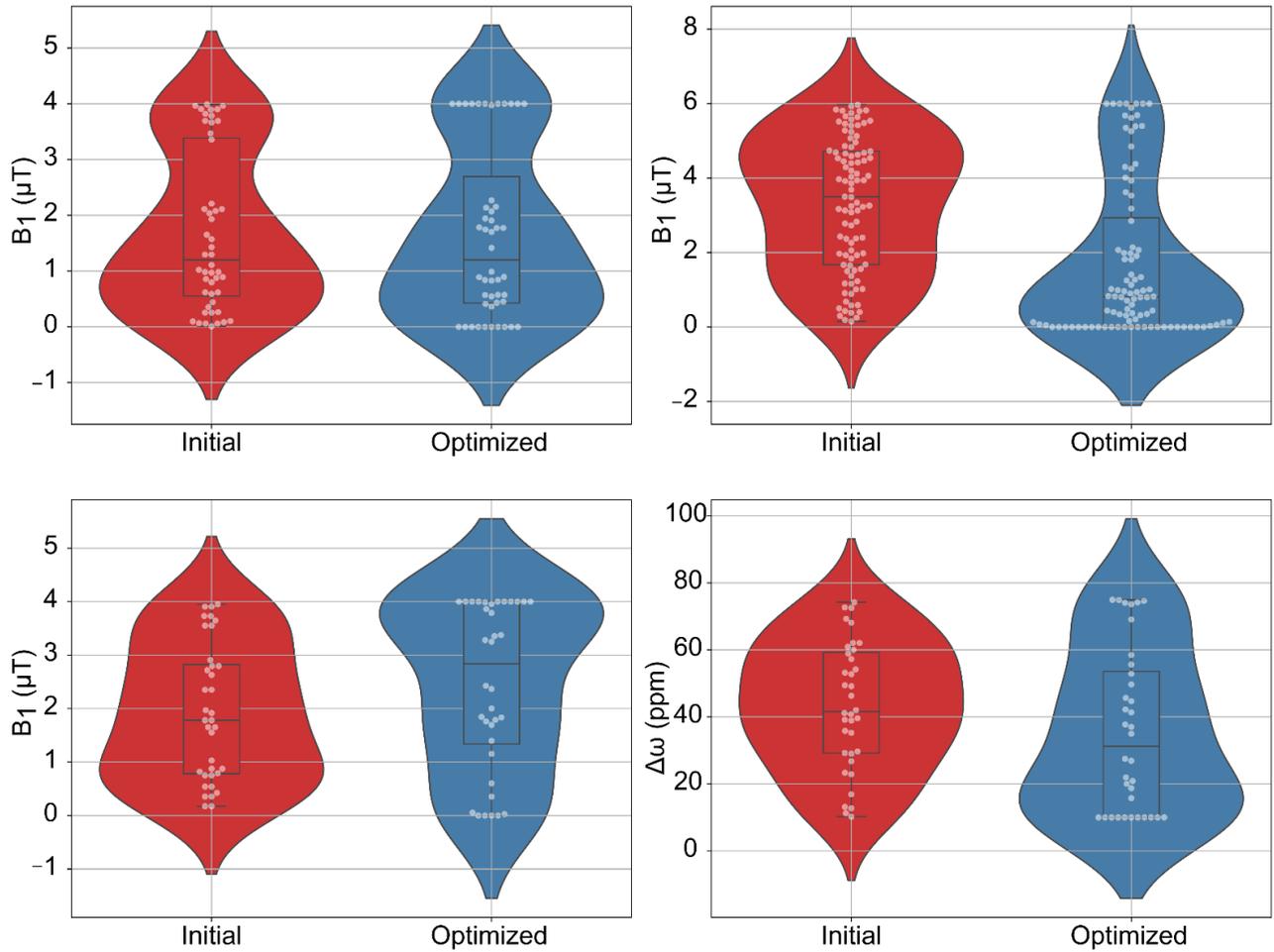

**Figure S11. Parameter distribution of MRF acquisition protocols: A comparison between randomly initialized and CRB optimized acquisition schedules. (Top left)** Distribution of saturation pulse powers used for CW phantom imaging[37]. **(Top right)** Distribution of saturation pulse powers used for PW phantom imaging[40]. **(Bottom left)** Distribution of saturation pulse powers, and **(Bottom right)** frequency offsets used for PW human imaging[40]. All other acquisition parameters are detailed in Section 2.4.

**Table S1. Dictionary properties.**

| Case | 7T CW L-arginine Imaging | 3T PW L-arginine Imaging | 3T In-vivo PW Semisolid MT Imaging |
|---|---|---|---|
| Water T1 (ms) | 2500:100:3300 | 2500:100:3300 | 800:100:3500 |
| Water T2 (ms) | 600:1200:50 | 400:50:2000 | 10:10:150 |
| T2* | - | 30 ms, 10 isochromats | |
| Solute T1 (ms) | Equal to water T1 | Equal to water T1 | |
| Solute T2 (ms) | 40 | 40 | - |
| Solute concentration (mM) | 10:5:120 | 10:5:120 | |
| Solute exchange rate (s$^{-1}$) | 100:10:1400 | 100:10:1000 | |
| Solute chemical shift (ppm) | 3 | | |
| Semisolid MT T1 (ms) | | | 1950 |
| Semisolid MT T2 (µs) | - | - | 40 |
| Semisolid MT proton volume fraction (%) | | | 1.8:1.8:27 |
| Semisolid MT proton exchange rate (s$^{-1}$) | | | 5:5:110 |
| Semisolid MT chemical shift (ppm) | | | -2 |
| Semisolid MT Lineshape | | | Lorentzian |
| Total number of entries | 665,783 | 792,792 | 129,360 |

**Table S2. Computational complexity timing**. In all measurements, the number of parallel workers for the dictionary generation was set to 16. An NVIDIA RTX3060 GPU was used for dot-product calculation.

| | Preclinical[a] | | Clinical L-arg[b] | | Clinical Brain[c] | |
|---|---|---|---|---|---|---|
| Acquisition protocol length (no. raw images) | 4 | 8 | 4 | 8 | 4 | 8 |
| Optimization time (hours) | 19.64±4.15 | 15.07±2.64 | 14.26±2.97 | 20.29±10.76 | 20.74±7.7 | 30.63±9.39 |
| Dictionary generation (sec) | 2.56±0.07 | 3.84±0.1 | 56.28±2.2 | 124.7±0.82 | 2.56±0.13 | 4.67±0.04 |
| Dictionary generation with T2* (sec) | | - | 587.7±6.2 | 1241±13.1 | 22.41±1.19 | 46.61±2.11 |
| GPU Dot-Product with overhead[d] (sec) | 1.13±0.07 | 2.16±0.1 | 1±0.08 | 1.16±0.27 | 11.59±0.05 | 11.99±0.09 |
| GPU Dot-Product (sec) | 0.28±0.02 | 0.98±0 | 0.23±0 | 0.36±0 | 10.61±0.17 | 10.98±0.05 |

[a]In the CW case, a dictionary with 665,783 entries was used

[b]In the PW L-arginine case, a dictionary with 406,406 entries was used.

[c]In the clinical MT case, a dictionary with 129,360 entries was used.

[d]Initialization overhead is roughly 1 second. In all cases, timing was performed for a single slice matching, while for whole brain human imaging, it was performed for all slices.